\documentstyle[amssymb,12pt,thmsa,sw20lart]{article}

\input{tcilatex}
\begin{document}

\title{Criteria of partial separability of multipartite qubit mixed-states}
\author{Zai-Zhe Zhong \\
Department of Physics, Liaoning Normal University, Dalian 116029, \\
Liaoning, China.}
\maketitle

\begin{abstract}
In this paper, we discuss the partial separability and its criteria problems
of multipartite qubit mixed-states. First we strictly define what is the
partial separability of a multipartite qubit system. Next we give a
reduction way from N-partite qubit density matrixes to bipartite qubit
density matrixes, and prove a necessary condition that a N-partite qubit
mixed-state to be partially separable is its reduction to satisfy the PPT
condition.

PACC numbers: 03.67.Mn; 03.65.Ud; 03.67.Hk
\end{abstract}

It is known that the entanglement problem is one of the most fascinating
features in modernistic quantum mechanics. Especially, it has recently been
recognized that entanglement is a very important resource in quantum
information processing, e.g. teleportation, quantum computation, quantum
cryptography and quantum communication, etc.. In the entanglement theory, an
important task is to find the criteria of separability of mixed-states. The
first important result is the well-known positive partial transposition
(PPT, Peres-Horodecki) criteria[1,2$]$ for $2\times 2$ and $2\times 3$
systems. There are many studies about the criteria of separability for the
multipartite systems, see [3-8$].$ About the problem of criteria of
separability of multipartite quantum systems, it is not completely solved as
yet.

Generally, the common so-called `separability', in fact, is the
`full-separability'. However, for multipartite systems the problems are more
complex: There yet is other concept of separability weaker than
full-separability, i.e. the `partial separability'$,$ e.g. the
A-BC-separability, B-AC-separability for a tripartite qubit pure-state $\rho
_{ABC}[8]$, etc.. Related to Bell-type inequalities and some criteria of
partial separability of multipartite systems, etc., see [9-12$].$

The main purpose of this paper is to discuss the partial separability and
the criteria problems of multipartite systems. First, we need to stricter
define the concept of partial separability, further we can find the simpler
criteria. Therefore, in the first part of this paper we discuss how to
define strictly the concept of the partial separability corresponding to a
partition. In the second part of this paper, we give a new way that an
arbitrary N-partite (N$\geq 3)$ qubit density matrix always can be reduced
in one step through to a bipartite qubit density matrix . In third part of
this paper, we prove an effective criterion: A necessary condition of a
N-partite qubit state to be partially separable\ with respect to a partition
is that the corresponding reduced bipartite qubit state in particular
satisfies\ the PPT condition. Some examples are given in the last part of
this paper.

Suppose that $\rho _{i_1i_2\cdots i_N}$ is a mixed-state for N-partite qubit
Hilbert space $H=\otimes _{s=1}^NH_s$ $,$ of which the standard basis is $%
\left\{ \otimes _{s=1}^N\mid i_s>\right\} (i_s=0,1).$ Let ${\Bbb Z}_N$ be
the integer set $\left\{ 1,2,\cdots ,N\right\} .$ If two subsets $\left(
r\right) _P\equiv \left\{ r_1,\cdots ,r_P\right\} $ and $\left( s\right)
_{N-P}\equiv \left\{ s_1,\cdots ,s_{N-P}\right\} $ in ${\Bbb Z}_N$ obey 
\begin{eqnarray}
\;1 &\leqslant &r_1<\cdots <r_P<N,\;1<s_1<\cdots <s_{N-P}\leqslant N 
\nonumber \\
\;\left( r\right) _P\cup \left( s\right) _{N-P} &=&{\Bbb Z}_N,\;\left(
r\right) _P\cap \left( s\right) _{N-P}=\emptyset \;
\end{eqnarray}
where $P$ is an integer, 1$\leqslant P\leqslant N-1,$ the pair $\left\{
\left( r\right) _P,\left( s\right) _{N-P}\right\} $ forms a partition of $%
{\Bbb Z}_N$, in the following we simply call it a `partition', and for the
sake of stress we denote it by symbol $\left( r\right) _P\Vert \left(
s\right) _{N-P}$. corresponds to a permutation $S_{\left( r\right) _P\Vert
\left( s\right) _{N-P}}\equiv \left( 
\begin{array}{cccccc}
1, & \cdots & P, & P+1, & \cdots & N \\ 
r_1, & \cdots & r_P, & s_1, & \cdots & s_{N-P}
\end{array}
\right) .$ By a partition $\left( r\right) _P\Vert \left( s\right) _{N-P}$,
a new matrix $\rho _{\left( r\right) _P\Vert \left( s\right) _{N-P}}$ from $%
\rho _{i_1i_2\cdots i_N}$ can be defined now, whose entries are determined
by 
\begin{equation}
\left[ \rho _{\left( r\right) _P\Vert \left( s\right) _{N-P}}\right]
_{j_1\cdots j_N,\;k_1\cdots k_N}=\left[ \rho \right] _{j_{r_1}\cdots
j_{r_P}j_{s_1}\cdots j_{s_{N-P}},\;k_{r_1}\cdots k_{r_P}k_{s_1}\cdots
k_{s_{N-P}}}
\end{equation}
For instance, $\rho _{A\Vert BCD}=\rho _{AB\Vert CD}\;=\rho _{ABC\Vert
D}=\rho _{ABCD}$, and $\left[ \rho _{AC\Vert BD}\right] _{ijkl,rstu}=\left[
\rho _{ABCD}\right] _{ikjl,\;rtsu},$ $\left[ \rho _{C\Vert ABD}\right]
_{ijkl,rstu}=\left[ \rho _{ABCD}\right] _{kijl,.trsu}$, etc.. Generally, $%
\rho _{\left( r\right) _P\Vert \left( s\right) _{N-P}}\neq \rho
_{i_1i_2\cdots i_N}$, unless $\left( r\right) _P\Vert \left( s\right) _{N-P}$
just maintains the natural order of ${\Bbb Z}_N$ (i.e. $\left( r\right)
_P=\left( 1,\cdots ,P\right) ,\left( s\right) _{N-P}=\left( P+1,\cdots
,N\right) )$, then $\rho _{\left( r\right) _P\Vert \left( s\right)
_{N-P}}=\rho _{i_1i_2\cdots i_N}.$

{\bf Lemma. }For any{\bf \ }partition $\left( r\right) _P\Vert \left(
s\right) _{N-P},$ $\rho _{\left( r\right) _P\Vert \left( s\right) _{N-P}}$
is still a N-partite qubit mixed-state.

{\bf Proof. }We only consider the case of tripartite qubit, the general
cases are completely similar (also see [11$]).$ Notice the permutation $%
S_{B\Vert AC}$ , then we have 
\begin{equation}
\;\rho _{B\Vert AC}=S\rho _{ABC}S^{\dagger },\;S=\left[ 
\begin{array}{cccccccc}
1 &  &  &  &  &  &  &  \\ 
& 1 &  &  &  &  &  &  \\ 
&  & 0 & 0 & 1 & 0 &  &  \\ 
&  & 0 & 0 & 0 & 1 &  &  \\ 
&  & 1 & 0 & 0 & 0 &  &  \\ 
&  & 0 & 1 & 0 & 0 &  &  \\ 
&  &  &  &  &  & 1 &  \\ 
&  &  &  &  &  &  & 1
\end{array}
\right]
\end{equation}
$S$ is an unitary matrix, therefore $\rho _{B\Vert AC}$ is still a
tripartite qubit mixed-state. As for the cases of $S_{A\Vert BC}$ and $%
S_{C\Vert AB},$ the lemma obviously holds. $\square $

Now, we consider how to more strictly define the partial separability.
Obviously, if a partition $\left( r\right) _P\Vert \left( s\right) _{N-P}$
maintains the natural order of ${\Bbb Z}_N$ (i.e. $\left( r\right) _P=\left(
1,2,\cdots ,P\right) ,$ $\left( s\right) _{N-P}=\left( P+1,P+2,\cdots
,N\right) ),$ then $\rho _{\left( r\right) _P\Vert \left( s\right)
_{N-P}}=\rho _{i_1i_2\cdots i_N}$ under the standard basis $\left\{ \otimes
_{s=1}^N\mid i_s>\right\} $, now the $\left( r\right) _P-\left( s\right)
_{N-P}$-separability can naturally be defined as that if $\rho
_{i_1i_2\cdots i_N}$ can be decomposed as $\rho _{\left( r\right) _P\Vert
\left( s\right) _{N-P}}=\rho _{i_1i_2\cdots i_N}=\sum\limits_\alpha p_\alpha
\rho _{\alpha ,\left( r\right) _P}\otimes \rho _{\alpha ,\left( s\right)
_{N-P}}$ with probabilities $p_\alpha $, where $\rho _{\alpha ,\left(
r\right) _P}$ and $\rho _{\alpha ,\left( s\right) _{N-P}}$, respectively,
are a $P$-partite and a $\left( N-P\right) $-partite qubit mixed-states of $%
\otimes _{m=1}^PH_m$ and $\otimes _{n=1}^{N-P}H_n$ for all $\alpha ,$ then
we call $\rho _{i_1i_2\cdots i_N}$ to be $\left( r\right) _P-\left( s\right)
_{N-P}$-separable$.$ However, if the natural order of ${\Bbb Z}_N$ has been
broken in $\left( r\right) _P\Vert \left( s\right) _{N-P}$ $($ i.e. $%
s_1<r_P),$ then generally $\rho _{\left( r\right) _P\Vert \left( s\right)
_{N-P}}\neq \rho _{i_1i_2\cdots i_N}$ , the case is different from the
above. For instance, we consider a normalized pure-state $\rho _{ABCD}=\mid
\Psi _{ABCD}><\Psi _{ABCD}\mid ,$ $\mid \Psi _{ABCD}>\in H_A\otimes
H_B\otimes H_C\otimes H_D$ of four spin-$\frac 12$ particles A, B, C and D.
Now, assume that $\mid \Psi _{ABCD}>$ has a special form as $\mid \Psi
_{ABCD}>=\sum\limits_{i,j,k,l=0,1}c_{ik}c_{jl}\mid i_A>\otimes \mid
j_B>\otimes \mid k_C>\otimes \mid l_D>,$ where $c_{ik},c_{jl}\in {\Bbb C}^1$%
. If we keep up to use the original standard basis$,$ then we cannot
directly see the partial separability, because this choice of basis is
unsuitable. If we choose other nature basis $\left\{ \mid i_A>\otimes \mid
k_C>\otimes \mid j_B>\otimes \mid l_D>\right\} $ $($ this, in fact, means
that we are using $\rho _{AC\Vert BD}),$ under which we can consider the
state $\mid \Psi _{ACBD}^{\prime }>=\mid \Psi _{AC}>\otimes \mid \Psi _{BD}>$
, where $\mid \Psi _{AC}>=\sum\limits_{i,k=0,1}c_{ik}\mid i_A>\otimes \mid
k_C>,$ $\mid \Psi _{BD}>=\sum\limits_{j,l=0,1}c_{jl}\mid j_B>\otimes \mid
l_D>.$ Now, $\rho _{AC\Vert BD}=\rho _{AC}\otimes \rho _{BD}$, where $,$ $%
\rho _{AC}=\mid \Psi _{AC}><\Psi _{AC}\mid ,\rho _{BD}=\mid \Psi _{BD}><\Psi
_{BD}\mid $. $\mid \Psi _{ABCD}>$ and $\mid \Psi _{ABCD}>$, in fact, are the
same in physics, therefore to call $\rho _{ABCD}$ AC-BD-separable is
completely reasonable. Similarly, for the rest. Generalize to the cases of
mixed-states, thus we can generally define the concept of partial
separability as follows.

{\bf Definition. }{\it For the partition} $\left( r\right) _P\Vert \left(
s\right) _{N-P}$ , {\it a N-partite qubit state }$\rho _{i_1i_2\cdots i_N}$ 
{\it of H}$=\otimes _{s=1}^NH_s$ {\it is} {\it called to be }$\left(
r\right) _P-\left( s\right) _{N-P}${\it -separable if the corresponding
state }$\rho _{\left( r\right) _P\Vert \left( s\right) _{N-P}}${\it \ can be
decomposed as } 
\begin{equation}
\rho _{\left( r\right) _P\Vert \left( s\right) _{N-P}}=\sum\limits_\alpha
p_\alpha \rho _{\alpha ,\left( r\right) _P}\otimes \rho _{\alpha ,\left(
s\right) _{N-P}}
\end{equation}
{\it where }$\rho _{\alpha ,\left( r\right) _P}${\it \ and }$\rho _{\alpha
,\left( s\right) _{N-P}}${\it , respectively,} {\it are a }$P${\it -partite
and a }$\left( N-P\right) ${\it -partite qubit state of }$\otimes
_{m=1}^PH_{r_m}${\it \ and }$\otimes _{n=1}^{N-P}H_{s_n}$ {\it for all }$%
\alpha ,$ {\it and }$0<p_\alpha \leq 1,\;\sum\limits_\alpha p_\alpha =1.$ 
{\it If }$\rho _{i_1i_2\cdots i_N}${\it \ is not }$\left( r\right) _P-\left(
s\right) _{N-P}${\it -separable, then we call it }$\left( r\right) _P-\left(
s\right) _{N-P}${\it -inseparable.}

For the distinct partitions $\rho _{i_1i_2\cdots i_N}$ can have distinct
separability. Of course, if a $\rho _{i_1i_2\cdots i_N}$ is partially
inseparable for some partition, then it must be entangled. Here, in passing,
we point out that how to find the general relations between the partial
separability and the ordinary separability (full-separability), generally,
is not a simple problem. For instance, there is such a multipartite qubit
mixed-state $\stackrel{\backsim }{\rho }$ $($see the theorem 1 and its proof
in [13,14]), $\stackrel{\backsim }{\rho }$ always is partially separable for
all possible partitions $\left( r\right) _P\Vert \left( s\right)
_{N-P}\left( 1\leqslant P\leqslant N-1\right) ,$ but $\stackrel{\backsim }{%
\rho }$ is entangled.

In order to find the criteria of \ partial separability, first we discuss
how to reduce a multipartite qubit density matrix in one step through to a
bipartite qubit density matrix. For a given partition $\left( r\right)
_P\Vert \left( s\right) _{N-P},$ let two sets $\left( r\right) _P$ and $%
\left( s\right) _{N-P},$ respectively, be separated anew as follows, 
\begin{eqnarray}
\left( r^{\prime }\right) _{P^{\prime }} &=&\left\{ r_1^{\prime },\cdots
,r_{P^{\prime }}^{\prime }\right\} ,\left( r^{\prime \prime }\right)
_{P^{\prime \prime }}=\left\{ r_1^{\prime \prime },\cdots ,r_{P^{^{\prime
\prime }}}^{\prime \prime }\right\} \text{ ( one of them can be the null set)%
}  \nonumber \\
\text{ }r_1^{\prime } &<&r_2^{\prime }<\cdots <r_{P^{\prime }}^{\prime
},\;\;r_1^{\prime \prime }<r_2^{\prime \prime }<\cdots \;<r_{P^{\prime
\prime }}^{\prime \prime }  \nonumber \\
\left( r\right) _P &=&\left( r^{\prime }\right) _{P^{\prime }}\cup \left(
r^{\prime \prime }\right) _{P^{\prime \prime }},\;\left( r^{\prime }\right)
_{P^{\prime }}\cap \left( r^{\prime \prime }\right) _{P^{\prime \prime
}}=\emptyset (0\leq P^{\prime },P^{\prime \prime }\leq P\text{ and }%
P^{\prime }+P^{\prime \prime }=P) \\
\left( s^{\prime }\right) _{Q^{\prime }} &=&\left\{ s_1^{\prime },\cdots
,s_{Q^{\prime }}^{\prime }\right\} ,\left( s^{\prime \prime }\right)
_{Q^{\prime \prime }}=\left\{ s_1^{\prime \prime },\cdots ,s_{Q^{\prime
\prime }}^{\prime \prime }\right\} \text{ ( one of them can be the null set)}
\nonumber \\
\;s_1^{\prime } &<&\text{ }s_2^{\prime }<\cdots <s_{Q^{\prime }}^{\prime
},\;\;s_1^{\prime \prime }<\text{ }s_2^{\prime \prime }<\cdots <s_{Q^{\prime
\prime }}^{\prime \prime }\text{ }  \nonumber \\
\;\left( s\right) _{N-P} &=&\left( s^{\prime }\right) _{Q^{\prime }}\cup
\left( s^{\prime \prime }\right) _{Q^{\prime \prime }},\left( s^{\prime
}\right) _{Q^{\prime }}\cap \left( s^{\prime \prime }\right) _{Q^{\prime
\prime }}=\emptyset \left( 0\leq Q^{\prime },Q^{\prime \prime }\leq N-P\text{
and }Q^{\prime }+Q^{\prime }=N-P\right)  \nonumber
\end{eqnarray}
now we rewrite the partition added these partitions as $\left[ \left(
r^{\prime }\right) _{P^{\prime }},\left( r^{\prime \prime }\right)
_{P^{\prime \prime }}\right] \Vert \left[ \left( s^{\prime }\right)
_{Q^{\prime }},\left( s^{\prime \prime }\right) _{Q^{\prime \prime }}\right] 
$. Now we define the matrix $\rho _{\left[ \left( r^{\prime }\right)
_{P^{\prime }},\left( r^{\prime \prime }\right) _{P^{\prime \prime }}\right]
\Vert \left[ \left( s^{\prime }\right) _{m-P^{\prime }},\left( s^{\prime
\prime }\right) _{m-P^{\prime \prime }}\right] }$ by 
\begin{eqnarray}
&&\rho _{\left[ \left( r^{\prime }\right) _{P^{\prime }},\left( r^{\prime
\prime }\right) _{P^{\prime \prime }}\right] \Vert \left[ \left( s^{\prime
}\right) _{m-P^{\prime }},\left( s^{\prime \prime }\right) _{m-P^{\prime
\prime }}\right] }\text{ = the submatrix in }\rho _{i_1\cdots i_N}\text{
consisting of all entries }  \nonumber \\
&&\text{with form as }\left[ \rho \right] _{x_1x_2\cdots x_N,\;y_1y_2\cdots
y_N}
\end{eqnarray}
which must be a 4$\times 4$ matrix, where the values of $x_k$ and $y_k\left(
k=1,\cdots ,N\right) ,$ respectively, are determined by 
\begin{eqnarray}
x_k &=&i\text{ for }k\in \left( r^{\prime }\right) _{P^{\prime }},\;x_k=1-i%
\text{ for }k\in \left( r^{\prime \prime }\right) _{P^{\prime \prime }} 
\nonumber \\
\;x_k &=&j\text{ for }k\in \left( s^{\prime }\right) _{Q^{\prime }},\;x_k=1-j%
\text{ for }k\in \left( s^{\prime \prime }\right) _{Q^{\prime \prime }} 
\nonumber \\
y_k &=&u\text{ for }k\in \left( r^{\prime }\right) _{P^{\prime }},\;y_k=1-u%
\text{ for }k\in \left( r^{\prime \prime }\right) _{P^{\prime \prime }} \\
\;y_k &=&v\text{ for }k\in \left( s^{\prime }\right) _{Q^{\prime }},\;y_k=1-v%
\text{ for }k\in \left( s^{\prime \prime }\right) _{Q^{\prime \prime }}\; 
\nonumber
\end{eqnarray}
where $i,j,u,v=0,1$. E.g. 
\begin{eqnarray}
\rho _{\left[ \left( AC\right) ,\emptyset \right] \Vert \left[ \left(
B\right) ,\left( D\right) \right] } &=&\text{the submatrix in }\rho _{ABCD}%
\text{ consisting of all entries with}  \nonumber \\
\text{ form as }\left[ \rho \right] _{iji\left( 1-j\right) ,uvu\left(
1-v\right) } &=&\left[ 
\begin{array}{llll}
\left[ \rho \right] _{0001,0001} & \left[ \rho \right] _{0001,0100} & \left[
\rho \right] _{0001,1011} & \left[ \rho \right] _{0001,1110} \\ 
\left[ \rho \right] _{0100,0001} & \left[ \rho \right] _{0100,0100} & \left[
\rho \right] _{0100,1011} & \left[ \rho \right] _{0100,1110} \\ 
\left[ \rho \right] _{1011,0001} & \left[ \rho \right] _{1011,0100} & \left[
\rho \right] _{1011,1011} & \left[ \rho \right] _{1011,1110} \\ 
\left[ \rho \right] _{1110,0001} & \left[ \rho \right] _{1110,0100} & \left[
\rho \right] _{1110,1011} & \left[ \rho \right] _{1110,1110}
\end{array}
\right]
\end{eqnarray}
etc.. Notice that there may be some repeated{\em \ }$\rho _{\left[ \left(
r^{\prime }\right) _{P^{\prime }},\left( r^{\prime \prime }\right)
_{P^{\prime \prime }}\right] \Vert \left[ \left( s^{\prime }\right)
_{m-P^{\prime }},\left( s^{\prime \prime }\right) _{m-P^{\prime \prime
}}\right] }$, e.g. {\em \ }$\rho _{\left[ \left( r^{\prime }\right)
_{P^{\prime }},\left( r^{\prime \prime }\right) _{P^{\prime \prime }}\right]
\Vert \left[ \left( s^{\prime }\right) _{m-P^{\prime }},\left( s^{\prime
\prime }\right) _{m-P^{\prime \prime }}\right] }$

$=\rho _{\left[ \left( r^{\prime }\right) _{P^{\prime }},\left( r^{\prime
\prime }\right) _{P^{\prime \prime }}\right] \Vert \left[ \left( s^{\prime
\prime }\right) _{m-P^{\prime \prime }},\left( s^{\prime }\right)
_{m-P^{\prime }}\right] }$

=$\rho _{\left[ \left( r^{\prime \prime }\right) _{P^{\prime \prime
}},\left( r^{\prime }\right) _{P^{\prime }}\right] \Vert \left[ \left(
s^{\prime }\right) _{m-P^{\prime }},\left( s^{\prime \prime }\right)
_{m-P^{\prime \prime }}\right] }${\em \ }$=\rho _{\left[ \left( r^{\prime
\prime }\right) _{P^{\prime \prime }},\left( r^{\prime }\right) _{P^{\prime
}}\right] \Vert \left[ \left( s^{\prime \prime }\right) _{m-P^{\prime \prime
}},\left( s^{\prime }\right) _{m-P^{\prime }}\right] },$ etc..

Now we define the 4$\times 4$ matrix {\it \ }$\rho _{\left( \left( r\right)
_P-\left( s\right) _{N-P}\right) }$ by 
\begin{eqnarray}
&&\rho _{\left( \left( r\right) _P-\left( s\right) _{N-P}\right) } \\
&=&\sum_{_{\text{ }\;}}\left( \text{not repeated }\rho _{\left[ \left(
r^{\prime }\right) _{P^{\prime }},\left( r^{\prime \prime }\right)
_{P^{\prime \prime }}\right] \Vert \left[ \left( s^{\prime }\right)
_{m-P^{\prime }},\left( s^{\prime \prime }\right) _{m-P^{\prime \prime
}}\right] }\text{ }\right)  \nonumber
\end{eqnarray}
where we take the sum for those possible $\left[ \left( r^{\prime }\right)
_{P^{\prime }},\left( r^{\prime \prime }\right) _{P^{\prime \prime }}\right]
\Vert \left[ \left( s^{\prime }\right) _{m-P^{\prime }},\left( s^{\prime
\prime }\right) _{m-P^{\prime \prime }}\right] $ that $\rho _{\left[ \left(
r^{\prime }\right) _{P^{\prime }},\left( r^{\prime \prime }\right)
_{P^{\prime \prime }}\right] \Vert \left[ \left( s^{\prime }\right)
_{m-P^{\prime }},\left( s^{\prime \prime }\right) _{m-P^{\prime \prime
}}\right] }$ are not repeated. For instance, we have (where for the set $%
\left( i_1,i_2,i_3\right) \equiv \left( A,B,C\right) $ we simply write $%
\left( A\right) \equiv \left( A\right) _1,\left( BC\right) \equiv \left(
BC\right) _2,$ and $\left( A-BC\right) \equiv \left( \left( A\right)
_1-\left( BC\right) _2\right) ,$ etc.) 
\begin{eqnarray}
\rho _{\left( A-BC\right) } &=&\rho _{\left[ \left( A\right) ,\emptyset
\right] \Vert \left[ \left( BC\right) ,\emptyset \right] }+\rho _{\left[
\left( A\right) ,\emptyset \right] \Vert \left[ \left( B\right) ,\left(
C\right) \right] }  \nonumber \\
\rho _{\left( B-ACD\right) } &=&\rho _{\left[ \left( B\right) ,\emptyset
\right] \Vert \left[ \left( ACD\right) ,\emptyset \right] }+\rho _{\left[
\left( B\right) ,\emptyset \right] \Vert \left[ \left( AC\right) ,\left(
D\right) \right] }+\rho _{\left[ \left( B\right) ,\emptyset \right] \Vert
\left[ \left( AD\right) ,\left( C\right) \right] }+\rho _{\left[ \left(
B\right) ,\emptyset \right] \Vert \left[ \left( A\right) ,\left( CD\right)
\right] }  \nonumber \\
\rho _{\left( AC-BD\right) } &=&\rho _{\left[ \left( AC\right) ,\emptyset
\right] \Vert \left[ \left( BD\right) ,\emptyset \right] }+\rho _{\left[
\left( AC\right) ,\emptyset \right] \Vert \left[ \left( B\right) ,\left(
D\right) \right] }+\rho _{\left[ \left( A\right) ,\left( C\right) \right]
\Vert \left[ \left( BD\right) ,\emptyset \right] }+\rho _{\left[ \left(
A\right) ,\left( C\right) \right] \Vert \left[ \left( B\right) ,\left(
D\right) \right] }  \nonumber \\
\rho _{\left( AC-BDE\right) } &=&\rho _{\left[ \left( AC\right) ,\emptyset
\right] \Vert \left[ \left( BDE\right) ,\emptyset \right] }+\rho _{\left[
\left( AC\right) ,\emptyset \right] \Vert \left[ \left( BD\right) ,\left(
E\right) \right] }+\rho _{\left[ \left( AC\right) ,\emptyset \right] \Vert
\left[ \left( BE\right) ,\left( D\right) \right] } \\
&&+\rho _{\left[ \left( AC\right) ,\emptyset \right] \Vert \left[ \left(
B\right) ,\left( DE\right) \right] }+\rho _{\left[ \left( A\right) ,\left(
C\right) \right] \Vert \left[ \left( BDE\right) ,\emptyset \right] }+\rho
_{\left[ \left( A\right) ,\left( C\right) \right] \Vert \left[ \left(
BD\right) ,\left( E\right) \right] }  \nonumber \\
&&+\rho _{\left[ \left( A\right) ,\left( C\right) \right] \Vert \left[
\left( BE\right) ,\left( D\right) \right] }+\rho _{\left[ \left( A\right)
,\left( C\right) \right] \Vert \left[ \left( B\right) ,\left( DE\right)
\right] }  \nonumber
\end{eqnarray}

In order to vividly describe the above reduction procedures, we see the
example from $\rho _{ABCD}$\ to $\rho _{\left( AC-BD\right) }$ . The process 
$\rho _{ABCD}\longrightarrow $\ $\rho _{\left( AC-BD\right) }=\rho _{\left[
\left( AC\right) ,\emptyset \right] \Vert \left[ \left( BD\right) ,\emptyset
\right] }+\rho _{\left[ \left( AC\right) ,\emptyset \right] \Vert \left[
\left( B\right) ,\left( D\right) \right] }+\rho _{\left[ \left( A\right)
,\left( C\right) \right] \Vert \left[ \left( BD\right) ,\emptyset \right]
}+\rho _{\left[ \left( A\right) ,\left( C\right) \right] \Vert \left[ \left(
B\right) ,\left( D\right) \right] }$ can be described as follows: We simply
read $\sigma _\vartriangle $ $\equiv \rho _{\left[ \left( AC\right)
,\emptyset \right] \Vert \left[ \left( BD\right) ,\emptyset \right] },\sigma
_{\times }\equiv $\ $\rho _{\left[ \left( AC\right) ,\emptyset \right] \Vert
\left[ \left( B\right) ,\left( D\right) \right] },\sigma _{*}\equiv \rho
_{\left[ \left( A\right) ,\left( C\right) \right] \Vert \left[ \left(
BD\right) ,\emptyset \right] },\sigma _{\wedge }\equiv \rho _{\left[ \left(
A\right) ,\left( C\right) \right] \Vert \left[ \left( B\right) ,\left(
D\right) \right] }$, then $\rho _{\left( AC-BD\right) }\equiv \sigma
_\vartriangle +\sigma _{\times }+\sigma _{*}+\sigma _{\wedge },$ where the
entries of the submatrixes $\sigma _\vartriangle ,\sigma _{\times },\sigma
_{\diamond }$ and $\sigma _{\wedge },$\ respectively, simply are represented
by `$\vartriangle $', `$\times $',`$*$' and `$\wedge $' (they all are some
entries of $\rho _{ABCD})$, i.e. 
\begin{eqnarray}
\sigma _\vartriangle &=&\left[ 
\begin{array}{llll}
\vartriangle & \vartriangle & \vartriangle & \vartriangle \\ 
\vartriangle & \vartriangle & \vartriangle & \vartriangle \\ 
\vartriangle & \vartriangle & \vartriangle & \vartriangle \\ 
\vartriangle & \vartriangle & \vartriangle & \vartriangle
\end{array}
\right] ,\sigma _{\times }=\left[ 
\begin{array}{llll}
\times & \times & \times & \times \\ 
\times & \times & \times & \times \\ 
\times & \times & \times & \times \\ 
\times & \times & \times & \times
\end{array}
\right]  \nonumber \\
\sigma _{*} &=&\left[ 
\begin{array}{llll}
\ast & * & * & * \\ 
\ast & * & * & * \\ 
\ast & * & * & * \\ 
\ast & * & * & *
\end{array}
\right] ,\sigma _{\wedge }=\left[ 
\begin{array}{llll}
\wedge & \wedge & \wedge & \wedge \\ 
\wedge & \wedge & \wedge & \wedge \\ 
\wedge & \wedge & \wedge & \wedge \\ 
\wedge & \wedge & \wedge & \wedge
\end{array}
\right]
\end{eqnarray}
in the 16$\times 16$ matrix $\rho _{ABCD},$ the distributions of four
submatrixes $\sigma _\vartriangle ,\sigma _{\times },\sigma _{*}$ , $\sigma
_{\wedge }$ are as in the following figure ($\sigma _{\times }$ is just the
matrix $\rho _{\left[ \left( AC\right) ,\emptyset \right] \Vert \left[
\left( B\right) ,\left( D\right) \right] }$ in Eq.(8)), which determines yet
four submatrixes $\sigma _\vartriangle ,\sigma _{\times },\sigma _{*}$ , $%
\sigma _{\wedge }$ 
\begin{equation}
\begin{array}{lllllllllllllllll}
& ^{_{_{0000}}} & ^{_{_{0001}}} & ^{_{_{0010}}} & ^{_{_{0011}}} & 
^{_{_{0100}}} & ^{_{_{0101}}} & ^{_{_{0110}}} & ^{_{_{0111}}} & ^{_{_{1000}}}
& ^{_{_{1001}}} & ^{_{_{1010}}} & ^{_{_{1011}}} & ^{_{_{1100}}} & 
^{_{_{1101}}} & ^{_{_{1110}}} & ^{_{_{^{_{1111}}}}} \\ 
^{_{_{0000}}} & \bigtriangleup &  &  &  &  & \bigtriangleup &  &  &  &  & 
\bigtriangleup &  &  &  &  & \bigtriangleup \\ 
^{_{_{0001}}} &  & \times &  &  & \times &  &  &  &  &  &  & \times &  &  & 
\times &  \\ 
_{^{_{0010}}} &  &  & * &  &  &  &  & * & * &  &  &  &  & * &  &  \\ 
_{^{_{0011}}} &  &  &  & \wedge &  &  & \wedge &  &  & \wedge &  &  & \wedge
&  &  &  \\ 
^{_{_{0100}}} &  & \times &  &  & \times &  &  &  &  &  &  & \times &  &  & 
\times &  \\ 
^{_{_{0101}}} & \bigtriangleup &  &  &  &  & \bigtriangleup &  &  &  &  & 
\bigtriangleup &  &  &  &  & \bigtriangleup \\ 
^{_{_{0110}}} &  &  &  & \wedge &  &  & \wedge &  &  & \wedge &  &  & \wedge
&  &  &  \\ 
_{^{_{0111}}} &  &  & * &  &  &  &  & * & * &  &  &  &  & * &  &  \\ 
^{_{_{1000}}} &  &  & * &  &  &  &  & * & * &  &  &  &  & * &  &  \\ 
_{^{_{1001}}} &  &  &  & \wedge &  &  & \wedge &  &  & \wedge &  &  & \wedge
&  &  &  \\ 
^{_{_{1010}}} & \bigtriangleup &  &  &  &  & \bigtriangleup &  &  &  &  & 
\bigtriangleup &  &  &  &  & \bigtriangleup \\ 
_{^{_{1011}}} &  & \times &  &  & \times &  &  &  &  &  &  & \times &  &  & 
\times &  \\ 
_{^{_{1100}}} &  &  &  & \wedge &  &  & \wedge &  &  & \wedge &  &  & \wedge
&  &  &  \\ 
^{_{_{1101}}} &  &  & * &  &  &  &  & * & * &  &  &  &  & * &  &  \\ 
^{_{_{1110}}} &  & \times &  &  &  &  &  &  &  &  &  & \times &  &  & \times
&  \\ 
^{_{_{1111}}} & \bigtriangleup &  &  &  &  & \bigtriangleup &  &  &  &  & 
\bigtriangleup &  &  &  &  & \bigtriangleup
\end{array}
\end{equation}

Similarly, we can consider higher dimensional cases. As for the ordinary
bipartite qubit state $\rho _{AB},$\ we can take $\rho _{\left( A-B\right)
}\equiv \rho _{AB}$.

Sum up, generally we can define the 4$\times 4$\ matrix $\rho _{\left(
\left( r\right) _P-\left( s\right) _{N-P}\right) }$\ for a given $\left(
r\right) _P\Vert \left( s\right) _{N-P}.$\ In addition,\ it is easily
verified that for any partition $\left( r\right) _P\Vert \left( s\right)
_{N-P},$\ $\rho _{\left( \left( s\right) _{N-P}-\left( r\right) _P\right) }$%
\ is just the transposition of $\rho _{\left( \left( u\right) _P-\left(
s\right) _{N-P}\right) },$\ therefore from viewpoint of partial
separability, we don't have to distinguish between the partitions $\left(
r\right) _P\Vert \left( s\right) _{N-P}$\ and $\left( s\right) _{N-P}\Vert
\left( r\right) _P.$

{\bf Theorem 1. }{\it For any partition }$\left( r\right) _P\Vert \left(
s\right) _{N-P},${\it \ }$\rho _{\left( \left( r\right) _P-\left( s\right)
_{N-P}\right) }${\it \ is a bipartite qubit mixed-state, therefore }$\rho
_{\left( \left( r\right) _P-\left( s\right) _{N-P}\right) }${\it , in fact,} 
{\it is a reduction of the N-partite qubit density matrix }$\rho
_{i_1i_2\cdots i_N}${\it . }

{\bf Proof. }The fact must proved only is that $\rho _{\left( \left(
r\right) _P-\left( s\right) _{N-P}\right) }$\ is surely a bipartite qubit
density matrix. Here we only discuss in detail the cases of quadripartite
qubit states, since the generalization is completely straightforward. In the
first place, we prove that the theorem holds for a pure-state $\rho _{ABCD}$%
. Suppose that $\rho _{ABCD}=\mid \Psi _{ABCD}><\Psi _{ABCD}\mid $ is a
normalized pure-state$,$ where $\;\mid \Psi
_{ABCD}>=\sum_{i,j,k,l=0,1}c_{ijkl}\mid i_A>\otimes \mid j_B>\otimes \mid
k_C>\otimes \mid l_D>,$ $\sum_{i,j,k,l=0,1}\left| c_{ijkl}\right| ^2=1$. Let 
\begin{eqnarray}
&\mid &\Phi _{\bigtriangleup }>=\sum_{i,j=0,1}c_{ijij}\mid i_x>\otimes \mid
j_y>,\;\mid \Phi _{\times }>=\sum_{i,j=0,1}c_{iji\left( 1-j\right) }\mid
i_x>\otimes \mid j_y> \\
&\mid &\Phi _{*}>=\sum_{i,j=0,1}c_{ij\left( 1-i\right) j}\mid i_x>\otimes
\mid j_y>,\;\mid \Phi _{\wedge }>=\sum_{i,j=0,1}c_{ij\left( 1-i\right)
j\left( 1-j\right) }\mid i_x>\otimes \mid j_y>  \nonumber
\end{eqnarray}
where $x$ and $y$ are two particles. Make normalization, we obtain $\mid
\varphi _{\bigtriangleup }>=\eta _{\bigtriangleup }^{-1}\mid \Phi
_{\bigtriangleup },\mid \varphi _{\times }>=\eta _{\times }^{-1}\mid \Phi
_{\times }>$,$\mid \varphi _{*}>=\eta _{*}^{-1}\mid \Phi _{*}>,\mid \varphi
_{\wedge }>=\eta _{\wedge }^{-1}\mid \Phi _{\wedge }>$ and $\rho
_{\bigtriangleup }=\mid \varphi _{\bigtriangleup }><\varphi _{\bigtriangleup
}\mid ,\;$\ $\rho _{\times }=\mid \varphi _{\times }><\varphi _{\times }\mid
,\;\rho _{*}=\mid \varphi _{*}><\varphi _{*}\mid ,$ $\rho _{\wedge }=\mid
\varphi _{\wedge }><\varphi _{\wedge }\mid ,$ where the normalization
factors are 
\begin{eqnarray}
\eta _{\bigtriangleup } &=&\sqrt{\sum_{i,j=0,1}\left| c_{ijij}\right| ^2}%
,\;\eta _{\times }=\sqrt{\sum_{i,j=0,1}\left| c_{iji\left( 1-j\right)
}\right| ^2}  \nonumber \\
\eta _{*} &=&\sqrt{\sum_{i,j=0,1}\left| c_{ij\left( 1-i\right) j}\right| ^2}%
,\;\eta _{\wedge }=\sqrt{\sum_{i,j=0,1}\left| c_{ij\left( 1-i\right) \left(
1-j\right) }\right| ^2}
\end{eqnarray}
It can be directly verified that from Eq.(10) we have 
\begin{equation}
\rho _{\left( AC-BD\right) }=\eta _{\bigtriangleup }^2\rho _{\bigtriangleup
}+\eta _{\times }^2\rho _{\times }+\eta _{*}^2\rho _{*}+\eta _{\wedge
}^2\rho _{\wedge }
\end{equation}
where $\rho _{\bigtriangleup },$\ $\rho _{\times },\rho _{*},\rho _{\wedge }$
all are bipartite qubit pure-states. It is easily seen that since $\mid \Psi
_{ABCD}>$ is normalized, $\eta _{\bigtriangleup }^2+\eta _{\times }^2+\eta
_{*}^2+\eta _{\wedge }^2=\sum\limits_{i,j,k,l=0,1}\left| c_{ijkl}\right|
^2=1 $. This means that $\rho _{\left( AC-BD\right) }$ is a bipartite qubit
mixed state for this pur-state $\rho _{ABCD}$.

Secondly, if $\rho _{ABCD}=\sum\limits_\alpha p_\alpha \rho _{\left( \alpha
\right) \left( ABCD\right) }$ is a mixed-state, where every $\rho _{\left(
\alpha \right) \left( ABCD\right) }$ is a quadripartite qubit pure-state
with probabilities $p_\alpha ,$ then from Eq.(10) we have $\rho _{\left(
AC-BD\right) }=\sum_\alpha p_{\alpha ,\;}\left( \rho _\alpha \right)
_{\left( AC-BD\right) }$ (where we simply read $\left( \rho _\alpha \right)
_{\left( AC-BD\right) }$ $\equiv \left( \rho _{\left( \alpha \right) \left(
ACBD\right) }\right) _{\left( AC-BD\right) }).$ Since every $\left( \rho
_\alpha \right) _{\left( AC-BD\right) }$ is a bipartite qubit mixed-state$,$ 
$\rho _{\left( AC-BD\right) }$ is a mixed-state.

A similar way can be extended to higher dimensional case, the key is that
when $\rho _{i_1,\cdots ,i_N}$ is a pure-state$,$ then $\rho _{\left[ \left(
r^{\prime }\right) _{P^{\prime }},\left( r^{\prime \prime }\right)
_{P^{\prime \prime }}\right] \Vert \left[ \left( s^{\prime }\right)
_{m-P^{\prime }},\left( s^{\prime \prime }\right) _{m-P^{\prime \prime
}}\right] }$

$=\mid \Psi _{\left[ \left( r^{\prime }\right) _{P^{\prime }},\left(
r^{\prime \prime }\right) _{P^{\prime \prime }}\right] \Vert \left[ \left(
s^{\prime }\right) _{m-P^{\prime }},\left( s^{\prime \prime }\right)
_{m-P^{\prime \prime }}\right] }><\Psi _{\left[ \left( r^{\prime }\right)
_{P^{\prime }},\left( r^{\prime \prime }\right) _{P^{\prime \prime }}\right]
\Vert \left[ \left( s^{\prime }\right) _{m-P^{\prime }},\left( s^{\prime
\prime }\right) _{m-P^{\prime \prime }}\right] },$where the pure-state $\mid
\Psi _{\left[ \left( r^{\prime }\right) _{P^{\prime }},\left( r^{\prime
\prime }\right) _{P^{\prime \prime }}\right] \Vert \left[ \left( s^{\prime
}\right) _{m-P^{\prime }},\left( s^{\prime \prime }\right) _{m-P^{\prime
\prime }}\right] }>$ id defined by 
\begin{eqnarray}
&\mid &\Psi _{\left[ \left( r^{\prime }\right) _{P^{\prime }},\left(
r^{\prime \prime }\right) _{P^{\prime \prime }}\right] \Vert \left[ \left(
s^{\prime }\right) _{m-P^{\prime }},\left( s^{\prime \prime }\right)
_{m-P^{\prime \prime }}\right] }>=\sum_{i,j=0,1}c_{x_1x_2\cdots x_N}\mid
x_1>\otimes \cdots \otimes \mid x_N>  \nonumber \\
&&\text{ (}x_1,x_2,\cdots ,x_N\text{ are determined by Eq.(}7\text{))}
\end{eqnarray}
Therefore we just have 
\begin{eqnarray}
&\mid &\Psi _{i_1,\cdots ,i_N}> \\
&=&\sum \left( \text{not repeated }\mid \Psi _{\left[ \left( r^{\prime
}\right) _{P^{\prime }},\left( r^{\prime \prime }\right) _{P^{\prime \prime
}}\right] \Vert \left[ \left( s^{\prime }\right) _{m-P^{\prime }},\left(
s^{\prime \prime }\right) _{m-P^{\prime \prime }}\right] }>\right)  \nonumber
\end{eqnarray}
where we take the sum for those possible $\left[ \left( r^{\prime }\right)
_{P^{\prime }},\left( r^{\prime \prime }\right) _{P^{\prime \prime }}\right]
\Vert \left[ \left( s^{\prime }\right) _{m-P^{\prime }},\left( s^{\prime
\prime }\right) _{m-P^{\prime \prime }}\right] $ that $\mid \Psi _{\left[
\left( r^{\prime }\right) _{P^{\prime }},\left( r^{\prime \prime }\right)
_{P^{\prime \prime }}\right] \Vert \left[ \left( s^{\prime }\right)
_{m-P^{\prime }},\left( s^{\prime \prime }\right) _{m-P^{\prime \prime
}}\right] }>$ are not repeated. By using of this relation, make the similar
states as in Eq.(13), and make generalization to mixes-states, we can prove
that generally, a mixed-state density matrix $\rho _{i_1\cdots i_N}$ can be
reduced through to the bipartite qubit density matrix $\rho _{\left( \left(
r\right) _P-\left( s\right) _{N-P}\right) }.$ $\square $

The following theorem is the main result in this paper, it is an application
of PPT condition for multipartite qubit systems.

{\bf Theorem 2 (Criterion).}{\it \ For a given partition }$\left( r\right)
_P\Vert \left( s\right) _{N-P}${\it , a necessary condition of a N-partite(N}%
$\geqslant 3)${\it \ qubit state }$\rho _{i_1i_2\cdots i_N}${\it \ to be }$%
\left( r\right) _P-\left( s\right) _{N-P}${\it -}$separable${\it \ is that
the reduced bipartite qubit mixed-state }$\rho _{\left( \left( r\right)
_P-\left( s\right) _{N-P}\right) \text{ }}$ {\it in particular} $satisfies$%
{\it \ the PPT condition.}

{\bf Proof. }We only discuss in detail the case of quadripartite qubit, it
can be straightforwardly generalized to the case of arbitrary N-partite
qubit. In the first place, we prove that this theorem holds for a
quadripartite qubit pure-state. Suppose that the pure-state $\rho _{ABCD}$
is AC-BD-separable. This means that if we choose the natural basis $\left\{
\mid i_A>\otimes \mid j_C>\otimes \mid r_B>\otimes \mid s_D>\right\} ,$ then 
$\rho _{AC\Vert BD}=\rho _{AC}\otimes \rho _{BD}$, where $\rho _{AC}=\mid
\Psi _{AC}><\Psi _{AC}\mid ,\;\mid \Psi
_{AC}>=\sum\limits_{i,j=0,1}c_{ij}\mid i_A>\otimes \mid
j_C>,\sum\limits_{i,j=0,1}\left| c_{ij}\right| ^2=1,$ and $\rho _{BD}=\mid
\Psi _{BD}><\Psi _{BD}\mid ,\;\mid \Psi
_{BD}>=\sum\limits_{r,s=0,1}d_{rs}\mid r_B>\otimes \mid s_D>$, $%
\sum\limits_{r,s=0,1}\left| d_{rs}\right| ^2=1.$ From the above ways, it
easily checked that the bipartite qubit mixed-state $\rho _{\left(
AC-BD\right) }$, in fact, can be rewritten as 
\begin{eqnarray}
\rho _{\left( AC-BD\right) } &=&\sigma _\vartriangle +\sigma _{\times
}+\sigma _{*}+\sigma _{\wedge }=\sigma _{\left( AC\right) }\otimes \sigma
_{\left( BD\right) }+\sigma _{\left( AC\right) }\otimes \sigma _{\left( B%
\stackrel{\vee }{D}\right) }  \nonumber \\
&&+\sigma _{\left( A\stackrel{\vee }{C}\right) }\otimes \sigma _{\left(
BD\right) }+\sigma _{\left( A\stackrel{\vee }{C}\right) }\otimes \sigma
_{\left( B\stackrel{\vee }{D}\right) }
\end{eqnarray}
where $\sigma _{\left( AC\right) }=\mid \Phi _{\left( AC\right) }><\Phi
_{\left( AC\right) }\mid $ , we already write $\mid \Phi _{\left( AC\right)
}>=\sum\limits_{i=0,1}e_i\mid i_x>,e_i\equiv c_{ii}$ and $\mid i_A>\otimes
\mid i_C>\longrightarrow \mid i_x>$. Similarly, $\sigma _{\left( A\stackrel{%
\vee }{C}\right) }=\mid \Phi _{\left( A\stackrel{\vee }{C}\right) }><\Phi
_{\left( A\stackrel{\vee }{C}\right) }\mid ,\ \mid \Phi _{\left( A\stackrel{%
\vee }{C}\right) }>=\sum\limits_{j=0,1}f_j\mid j_x>$, $f_j\equiv c_{j\left(
1-j\right) }$ and $\mid j_B>\otimes \mid \left( 1-j\right)
_D>\longrightarrow \mid j_x>$, and similarly for $\sigma _{\left( BD\right)
},\sigma _{\left( B\stackrel{\vee }{D}\right) },$ etc.. Now, $\rho _{\left(
AC-BD\right) }$ can be written as 
\begin{eqnarray}
\rho _{\left( AC-BD\right) } &=&\eta _{\left( AC\right) }^2\eta _{\left(
BD\right) }^2\rho _{\left( AC\right) }\otimes \rho _{\left( BD\right) }+\eta
_{\left( AC\right) }^2\eta _{\left( B\stackrel{\vee }{D}\right) }^2\rho
_{\left( AC\right) }\otimes \rho _{\left( B\stackrel{\vee }{D}\right) } \\
&&+\eta _{\left( A\stackrel{\vee }{C}\right) }^2\eta _{\left( BD\right)
}^2\rho _{\left( A\stackrel{\vee }{C}\right) }\otimes \rho _{\left(
BD\right) }+\eta _{\left( A\stackrel{\vee }{C}\right) }^2\eta _{\left( B%
\stackrel{\vee }{D}\right) }^2\rho _{\left( A\stackrel{\vee }{C}\right)
}\otimes \rho _{\left( B\stackrel{\vee }{D}\right) }  \nonumber
\end{eqnarray}
where $\rho _{\left( AC\right) }=\left( \eta _{\left( AC\right) }\right)
^{-1}\mid \Phi _{\left( AC\right) }><\Phi _{\left( AC\right) }\mid ,\eta
_{\left( AC\right) }=\sqrt{\sum\limits_{i=0,1}\left| c_{ii}\right| ^2}.$
Now, $\rho _{\left( AC\right) }$ is a state of a single particle. Similarly,
for $\rho _{\left( A\stackrel{\vee }{C}\right) },\rho _{\left( BD\right)
},\rho _{\left( B\stackrel{\vee }{D}\right) }.$ Since 
\begin{eqnarray}
&&\eta _{\left( AC\right) }^2\eta _{\left( BD\right) }^2+\eta _{\left(
AC\right) }^2\eta _{\left( B\stackrel{\vee }{D}\right) }^2+\eta _{\left( A%
\stackrel{\vee }{C}\right) }^2\eta _{\left( BD\right) }^2+\eta _{\left( A%
\stackrel{\vee }{C}\right) }^2\eta _{\left( B\stackrel{\vee }{D}\right) }^2 
\nonumber \\
&=&\left( \eta _{\left( AC\right) }^2+\eta _{\left( A\stackrel{\vee }{C}%
\right) }^2\right) \left( \eta _{\left( BD\right) }^2+\eta _{\left( B%
\stackrel{\vee }{D}\right) }^2\right) =1\times 1=1
\end{eqnarray}
therefore $\rho _{\left( AC-BD\right) }$ is a separable bipartite qubit
mixed-state. The PPT condition for separability of 2$\times 2$ systems is
sufficient and necessary[2], thus $\rho _{\left( AC-BD\right) }$ satisfies\
the PPT condition. Similarly, for other partial separability.

Secondly, we prove that this theorem holds yet for partially separable
mixed-states. Suppose that $\rho _{ABCD}$ is a AC-BD-separable mixed-state,
then under the same natural basis there is a decomposition as $\rho
_{AB\Vert CD}=\sum_\alpha p_\alpha \rho _{\left( \alpha \right) \left(
AC\right) }\otimes \rho _{\left( \alpha \right) \left( BD\right) }$, where $%
\rho _{\left( \alpha \right) \left( AC\right) }$ and $\rho _{\left( \alpha
\right) \left( BD\right) }$ both are bipartite qubit pure-states as in the
above for all $\alpha ,$ $0<p_\alpha \leq 1,$ $\sum\limits_\alpha p_\alpha
=1.$ From the above reduction operation, obviously we have 
\begin{equation}
\rho _{\left( AC-BD\right) }=\sum_\alpha p_\alpha \left[ \rho _{\left(
\alpha \right) \left( AC\right) }\otimes \rho _{\left( \alpha \right) \left(
BD\right) }\right] _{\left( AC-BD\right) }
\end{equation}
According to the above mention, every $\left[ \rho _{\left( \alpha \right)
\left( AC\right) }\otimes \rho _{\left( \alpha \right) \left( BD\right)
}\right] _{\left( AC-BD\right) }$ is a separable bipartite qubit
mixed-state, this leads to that the convex sum $\rho _{\left( AC-BD\right) }$
in Eq.(21) still is a separable bipartite qubit mixed-state, and it must
satisfy\ the PPT condition.

Similarly, we cane prove higher dimensional cases. $\square $

{\bf Corollary. }{\it If the reduced bipartite\ qubit mixed-state }$\left(
\rho _{i_1i_2\cdots i_N}\right) _{\left( \left( r\right) _P-\left( s\right)
_{N-P}\right) }${\it \ violates the PPT condition for a partition }$\left(
r\right) _P\Vert \left( s\right) _{N-P}${\it , then }$\rho _{i_1i_2\cdots
i_N}${\it \ is }$\left( r\right) _P-\left( s\right) _{N-P}${\it -inseparable
and} {\it entangled.}

It, in fact, is the inverse-negative proposition of Theorem 2.

{\bf Examples. }Consider two{\bf \ }tripartite qubit states 
\begin{eqnarray}
\rho _{ABC}^{\prime } &=&\left[ 
\begin{array}{cccccccc}
0 &  &  &  &  &  &  &  \\ 
& \frac{1-x}4 &  &  &  &  &  &  \\ 
&  & \frac{1-x}4 &  &  &  &  &  \\ 
&  &  & \frac x2 & -\frac x2 &  &  &  \\ 
&  &  & -\frac x2 & \frac x2 &  &  &  \\ 
&  &  &  &  & \frac{1-x}4 &  &  \\ 
&  &  &  &  &  & \frac{1-x}4 &  \\ 
&  &  &  &  &  &  & 0
\end{array}
\right]  \nonumber \\
\rho _{ABC}^{\prime \prime } &=&\left[ 
\begin{array}{cccccccc}
0 &  &  &  &  &  &  &  \\ 
& \frac{1-x}4 &  &  &  &  &  &  \\ 
&  & \frac x2 & 0 & 0 & -\frac x2 &  &  \\ 
&  & 0 & \frac{1-x}4 & 0 & 0 &  &  \\ 
&  & 0 & 0 & \frac{1-x}4 & 0 &  &  \\ 
&  & -\frac x2 & 0 & 0 & \frac x2 &  &  \\ 
&  &  &  &  &  & \frac{1-x}4 &  \\ 
&  &  &  &  &  &  & 0
\end{array}
\right]
\end{eqnarray}
then we have 
\begin{equation}
\left( \rho _{ABC}^{\prime }\right) _{\left( A-BC\right) }=\left( \;\rho
_{ABC}^{\prime \prime }\right) _{\left( B-AC\right) }=\rho _W
\end{equation}
where $\rho _W$ is the Werner state[1,15$]$ which consists of a singlet
fraction $x$ and a random fraction $(1-x)$, 
\begin{eqnarray}
\left[ \rho _W\right] _{ij,rs} &=&xS_{ij,rs}+\frac 14\left( 1-x\right)
\delta _{ir}\delta _{js}  \nonumber \\
S_{01,01} &=&S_{10,10}=-S_{01,10}=-S_{10,01}=\frac 12 \\
&&\text{and all the other components of }S\text{ vanish}.  \nonumber
\end{eqnarray}
It is known[1$]$ that when $\frac 13<x\leq 1$ $\rho _W$ violates the PPT
condition, it leads to that $\rho _{ABC}^{^{\prime }}$ is A-BC-inseparable
and $\rho _{ABC}^{\prime \prime }$ is B-AC-inseparable.

By using of the above theorems and corollary, in some special cases we can
make a $N$-partite qubit from $2^{N-2}$ bipartite qubit states, which is
partially inseparable for a given partition. As in the above, for the case
of tripartite qubit we take two bipartite qubit states $\sigma _{\left(
1\right) },$ $\sigma _{\left( 2\right) }$ and real numbers $p_1,$ $%
p_2,\;0<p_1,$ $p_2\leq 1$ such that $\sigma =p_1\sigma _{\left( 1\right) }+$ 
$p_2$ $\sigma _{\left( 2\right) }$ is a bipartite qubit entangled state (
then it violates the PPT condition). If we want to construct a tripartite
qubit entangled state $\rho _{ABC}$ which is B-AC-inseparable, then we can
take the entries of $\rho _{ABC}$ by 
\begin{eqnarray}
\left[ \rho _{ABC}\right] _{ijk,rst} &=&p_1\left[ \sigma _{\left( 1\right)
}\right] _{ji,sr},\text{ for }k=i\text{ and }t=r\;  \nonumber \\
\left[ \rho _{ABC}\right] _{ijk,rst} &=&p_2\left[ \sigma _{\left( 2\right)
}\right] _{ji,sr},\text{ for }k=1-i\text{ and }t=1-r \\
\left[ \rho _{ABC}\right] _{ijk,rst} &=&0,\text{ for the rest }%
(i,j,k,r,s,t=0,1)  \nonumber
\end{eqnarray}
It can be verified that $\rho _{ABC}$ is a tripartite qubit mixed-states,
and is B-AC-inseparable. In fact, $\left( \rho _{ABC}\right) _{\left(
B-AC\right) }=\tau $ which violates the PPT condition. Similarly, for A-BC
and C-AB. The above way can be generalized to obtain a $\left( r\right)
_P-\left( s\right) _{N-P}$-inseparable state from a bipartite qubit
entangled mixed-state in form as $\tau =\sum\limits_{i=1}^{2^{N-2}}p_i\sigma
_{\left( i\right) },$ where all $\sigma _{\left( i\right) }$ are some
bipartite qubit states.

\_\_\_\_\_\_\_\_\_\_\_\_\_\_\_\_\_\_\_\_\_\_\_\_\_\_\_\_\_\_\_\_\_\_\_\_\_\_%
\_\_\_\_\_\_\_\_\_\_\_\_\_\_\_\_\_\_\_\_\_\_\_\_\_\_\_\_\_\_\_\_\_\_\_\_\_\_%
\_\_\_\_\_-

\bigskip

\bigskip

{\Large References}

[1] A. Peres, Phys. Rev. Lett., {\bf 77}(1996)1413.

[2] M. Horodecki, P. Horodecki, and R. Horodecki, Phys. Lett. A,

\ \ \ \ {\bf 223}(1996)1.

[3] S. Wu. X. Chen, and Y. Zhang, Phys. Lett. A, {\bf 275}(2000)244.

[4] M. Horodecki, P. Horodecki, and R. Horodecki,

\ \ \ \ quant-ph/0006071.

[5] B. M. Terhal, J. Theo. Compu. Sci., {\bf 287}(1)(2002)313.

[6] M. Horodecki, P. Horodecki, and R. Horodecki,

\ \ \ \ quant-ph/0206008.

[7] K. Chen, and L. A. Wu, Phys. Lett. A, {\bf 306}(2002)14.

[8] W. D\"{u}r, G. Vidal, and J. I. Cirac, Phys. Rev. A, {\bf 62}%
(2000)062314.

[9] M. Seevinck and G. Svetlichny, Phys. Rev. Lett., {\bf 89}(2002)060401.

[10] A. O. Pittenger and M. H. Rubin, Phys. Rev. A, {\bf 62}(2000)032313.

[11] A. M. Wang, quant-ph/0305016.

[12] T. Yamakami, quant-ph/0308072.

[13] C. H. Bennett , D. P. DiVincenzo , T. Mor, P. W. Shor,

\ \ \ \ \ J. A. Smolin, and B. M. Terhal, Phys. Rev. Lett., {\bf 82}%
(1999)5385.

[14] D. P. DiVincenzo , T. Mor, P. W. Shor, J. A. Smolin,

\ \ \ \ \ and B. M. Terhal, Comm. Math. Phys., {\bf 238}(2003)379.

[15] J. Blank and P. Exner, Acta Univ. Carolinae, Math. Phys. 18(1977)3.

\end{document}